\documentclass[twocolumn,trackchanges]{aastex7}
\usepackage{booktabs}
\usepackage{longtable}
\usepackage{array}
\usepackage{amsmath}
\usepackage{graphicx}
\usepackage{hyperref}
\hypersetup{linkcolor=red,citecolor=blue,filecolor=cyan,urlcolor=magenta}

\newcommand{\tefft}{\ensuremath{T_{\mathrm{eff}}}}
\usepackage{hyperref}

\begin{document}


\title{Emergence of a lithium dip in $\sim$35\,Myr ``Snake" Open Clusters}

\author{Yun-Yi Zhang}
\affiliation{School of Science, Hangzhou Dianzi University, Hangzhou 310018, China} 
\affiliation{Zhejiang Branch of National Astronomical Data Center, Hangzhou 310018, China} 
\email{}

\author{Hai-Jun Tian}
\affiliation{School of Science, Hangzhou Dianzi University, Hangzhou 310018, China} 
\affiliation{Zhejiang Branch of National Astronomical Data Center, Hangzhou 310018, China}
\email[show]{hjtian@hdu.edu.cn}

\author{Jian-Rong Shi}
\affiliation{National Astronomical Observatories, Chinese Academy of Sciences, Beijing 100101}
\affiliation{University of Chinese Academy of Sciences, Beijing 100049, People's Republic of China}
\email[show]{sjr@bao.ac.cn}
\author{Cheng-Cheng Xie} 
\affiliation{School of Science, Hangzhou Dianzi University, Hangzhou 310018, China} 
\affiliation{Zhejiang Branch of National Astronomical Data Center, Hangzhou 310018, China}
\email[show]{ccx@hdu.edu.cn}
\author{Xiang-Ming Yang}
\affiliation{School of Science, Hangzhou Dianzi University, Hangzhou 310018, China} 
\affiliation{Zhejiang Branch of National Astronomical Data Center, Hangzhou 310018, China}
\email{}


\begin{abstract}
    
 We report the discovery of a lithium dip (Li-dip) in the stellar ``Snake'' (age = $35 \pm 5$\,Myr), challenging the classical view that Li-dips emerge only at ages $\gtrsim 150$\,Myr. Using high-resolution spectra from \textsc{galah} DR4 ($R \sim 28,000$) for 211 member stars, we identify a clear depletion feature in a \tefft\ range of 6200--6800\,K with a depth of $\Delta A(\mathrm{Li}) \approx 0.40$\,dex. Our analysis reveals two key advances: the Li-dip appears $\gtrsim$100\,Myr earlier than the previous observations, and within the dip temperature range, a significant correlation is found between rotational velocity and lithium depletion. Specifically, fast rotators ($v \sin i > 25$~km~s$^{-1}$) exhibit stronger lithium depletion than slow rotators ($v \sin i < 25$\,km\,s$^{-1}$). This trend suggests that faster rotators develop stronger rotational shear at the convective–radiative boundary, which enhances turbulent mixing and accelerates lithium destruction. It is also found that the lower temperature edge of the lithium plateau can reach as low as 5500\,K for the young open clusters. 
\end{abstract}



\section{INTRODUCTION}
\label{sec:intro}  

In stellar interiors, lithium (Li) is destroyed by proton-capture reactions at temperatures exceeding $2.5 \times 10^6$\,K, and therefore survives only in the outermost layers of stars \citep{deliyannis1991lithium}. According to standard stellar evolution theory, Li depletion occurs primarily at the base of the surface convection zone during the pre-main sequence phase for G dwarfs \citep{Iben1967, Deliyannis1990, Cummings_2017}. The theory predicts that significant Li depletion takes place as stars develop deep convective envelopes, which transport surface material to hotter interior regions where Li is burned.

According to standard stellar evolution models, for stars with temperatures between 6200 and 6800\,K, the base of their convective zone is not hot enough 
for lithium fusion reactions; thus, the models predict that no lithium dip (Li-dip)  should appear \citep{Deliyannis_1997}. Observations indicated that the open clusters younger than 150\,Myr do not show this Li-dip, such as NGC\,2516 \citep[100\,Myr;][]{Jeffries1998}, Pleiades \citep[125\,Myr;][]{Bouvier2018}), M\,35 \citep[150\,Myr;][]{Jeffries2021}  while, it is obvious in intermediate-age and old open clusters, e.g., NGC\,3532 \citep[400\,Myr;][]{Steinhauer2025}, M\,48 \citep[420\,Myr;][]{Sun_2023}, Hyades and Praesepe \citep[650\,Myr;][]{Balachandran1995TheLD,Cummings_2017}, and 
NGC\,2420 \cite[1.7\,Gyr;][]{Semenova2020}, NGC\,3680 \citep[1.75\,Gyr;][]{2009AJ....138.1171A}, NGC\,2506 \citep[1.85\,Gyr;][]{Anthony-Twarog2018}, NGC\,752 \citep[1.9\,Gyr;][]{Hobbs1986,Balachandran1995TheLD}, NGC\,2243 \citep[3.6\,Gyr;][]{Anthony_Twarog_2021},  M\,67 \citep[3.9\,Gyr;][]{Balachandran1995TheLD}, and NGC\,188 \citep[6.3\,Gyr;][]{Sun_2025}.

The discrepancy between the model predictions and observed Li abundances suggests that additional physical processes beyond the standard theory contribute to Li depletion. Several non-standard mechanisms have been proposed, including mass loss \citep{Deliyannis1990,Sun_2025}, microscopic diffusion \citep{Richer_1993}, rotationally induced mixing \citep{Pinsonneault_1989}, and meridional-circulation-driven radial mixing \citep{Li2025ApJ...991..149L}.

Here, we report the discovery of Li-dip in our previously reported young stellar Snake structure \citep[age $\sim$ $35 \pm 5$\,Myr;][]{Tian2020,Wang2022,Yang_2024} in the solar neighborhood (distance $\approx 300-400$\,pc). This paper is organized as follows. In Section~\ref{sec:data}, the criteria for sample selection are described. Section~\ref{Lithium Abundance} explains the determination of Li abundances using the spectral synthesis method based on high-resolution spectra and validation of our results. Section~\ref{result} presents the results of our measurements. In Section~\ref{sec:discussion}, we analyze and interpret these results. Finally, the conclusions are shown in the last Section.

\begin{figure}
    \centering
\includegraphics[width=1\linewidth]{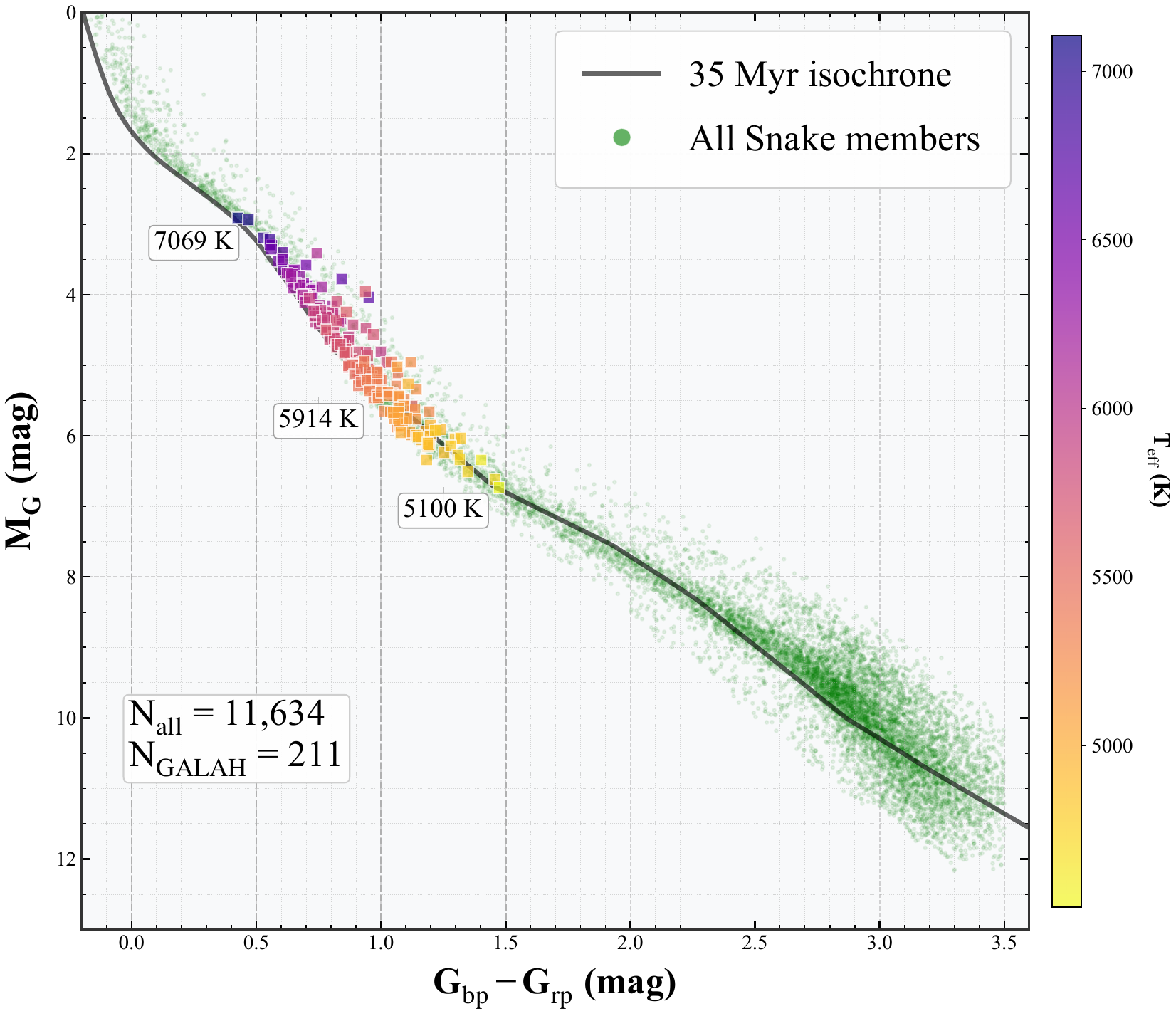}
\caption{The best-fit 35\,Myr isochrone for our stellar sample is displayed in the figure, with the 11,634 candidate member stars of the Snake highlighted in green. Among these, the 211 sources cross-matched with \textsc{galah} DR4 are marked by squares, which are colored according to their \tefft. The mean values in three consecutive bins of $G_{\rm bp}$ - $G_{\rm rp}$, spanning 0.0 to 1.5\,mag in steps of 0.5\,mag, are 7069\,K, 5914\,K, and 5100\,K, respectively.
}
\label{fig:fig1}
\end{figure}

\section{Sample stars and spectra} 
\label{sec:data}

\subsection{Initial Sample of the Snake Member Candidates} 
The giant Snake structure exhibits a coherent spatial distribution that extends over 200\,pc. The initial membership contains 1980 stars \citep{Tian2020}, and then expands to 11,746 candidates with 13 embedded open clusters \citep{Wang2022} through the analysis of \textit{Gaia} DR3 data. Its members include intermediate-mass and low-mass pre-main sequence (PMS) stars, extending across hundreds of parsecs \citep{Yang_2024}. The Snake exhibits three defining characteristics that establish it as an exceptional laboratory for stellar evolution studies: (1) It constitutes a single stellar population, with members aligning precisely along a single isochrone in the color-absolute magnitude diagram (CMD), indicating coeval formation $35 \pm 5$\,Myr ago \citep[see Figure~\ref{fig:fig1};][] {Tian2020}; (2) It displays a continuous mass function with a mass range of $0.5$ to $2\,M_\odot$ without discontinuities \citep{Yang_2024}, enabling comprehensive studies of mass-dependent processes; and (3) It maintains remarkable chemical homogeneity, showing a uniform abundance distribution with a mean metallicity of [Fe/H] $\approx 0.055 \pm 0.051$\,dex \citep[][]{Wang2022} given by \textsc{galah} DR3, consistent with formation from a well-mixed molecular cloud.
These characteristics collectively enable unprecedented investigations of coeval stellar formation, PMS evolution, and environmental effects. 
The youth and spatial coherence of Snake provide particularly stringent constraints on angular momentum transport mechanisms in low-mass stars, while its chemical uniformity offers a pristine testbed for studying metallicity-dependent processes in stellar evolution.

\subsection{Spectra from \textsc{galah} DR4} \label{subsec:data-source}

The high-resolution spectra ($R \sim 28,000$) of the sample stars were obtained from the \textsc{galah}  Data Release 4 (DR4) public archive \citep{Buder2025PASA}. The survey provides uniformly derived stellar parameters, such as effective temperature ($T_{\rm eff}$), surface gravity (log\,$g$), metallicity ([Fe/H]), micro-turbulent velocity ($\xi_{\rm t}$), Li abundances, and rotational velocity ($v$\,sin\,$i$) for each star.

Through cross-matching the stars in the Snake with those of \textsc{galah} DR4 \citep{Buder2025PASA}, we obtained a sample of 321 members confirmed by both kinematic and chemical criteria. Since the lithium line at 6708\,\AA\ corresponds to \texttt{ccd3} in \textsc{galah} DR4, we applied the following selection criteria to obtain a sample for reliable lithium abundance analysis.
\begin{itemize}
    \item Spectral quality and temperature range: Sources with high-resolution spectra having a signal-to-noise ratio (\texttt{snr\_px\_ccd3}) greater than 50 in the relevant \texttt{ccd3} with an effective temperature range of 4500\,K to 7200\,K. 
    \item Metallicity outlier removal: A 3$\sigma$ clipping was applied to the metallicity ([Fe/H]) values to exclude outliers. 
\end{itemize}

To ensure the inclusion of all stars with measured $v\sin i$, preserving the full range of rotational velocities for our analysis, we included all the member stars. The final sample contains 211 stars for subsequent analysis. It should be pointed out that the star \texttt{180129003601013} was removed because of its anomalous emission-like feature around 6708\,\AA, although its  signal-to-noise ratio is high. 

Given the different analysis pipelines used in \textsc{GALAH} DR3 and DR4, we performed a consistency check on the key stellar parameters for our 211 sample stars between the two releases. Our comparison shows that the \textit{A}(Li) values in DR4 are systematically lower than those in DR3, whereas no significant differences are found for $T_{\mathrm{eff}}$ or $v \sin i$ (except for stars with $v \sin i > 40$\,km\,s$^{-1}$). In addition, \textit{A}(Li) is not available for some stars in either data release. For these reasons, the lithium abundances presented in the remainder of this work are derived from our own procedure, as described in Section~\ref{Lithium Abundance}. Other key parameters, such as $v \sin i$ and $T_{\mathrm{eff}}$, are adopted directly from the \textsc{GALAH} DR4 catalog.

\section{Lithium abundances} \label{Lithium Abundance}

Although \textsc{galah} DR4 provided a valuable resource of homogeneously derived stellar parameters and elemental abundances, including lithium \citep{Buder2025PASA}, we would like to perform an independent measurement. As noted by \citet{Buder2025PASA}, a decrease in accuracy in abundance measurements was found, such as Eu. The differences between \textsc{galah} DR3 and DR4 are likely caused by the substantial change in the methodology between these two releases. In particular, \textsc{galah} DR3  derived abundances using the spectral synthesis method \citep{Buder2021} with Spectroscopy Made Easy \citep[][]{piskunov2017}, while the \textsc{galah} DR4's abundances \citep{Buder2025PASA} were determined with a neural network trained on synthetic spectra \citep{kane2025}. To derive robust and consistent lithium abundances, we performed a dedicated spectral re-analysis to measure \textit{A}(Li) for all stars in our sample.

\subsection{The determination of Li Abundance} 

For the sample members, the abundance of lithium ({\it A}(Li)\footnote{{\it A}(Li)= log$_{10}$(n$_{\rm {Li}}$/n$_{\rm {H}}$) + 12, where $n_{\rm {Li}}$ and $n_{\rm {H}}$ are the number densities of lithium and hydrogen atoms, respectively.}) were determined from the \ion{Li}{1} resonance line at 6708\,{\AA} under both the local thermodynamic equilibrium (LTE) and non-local thermodynamic equilibrium (NLTE) assumptions using the spectral synthesis method. We derived the lithium abundances using the relation of {\it A}(Li) = [Li/Fe] + [Fe/H] + {\it A}(Li)$_{\odot}$ with the solar lithium abundance of {\it A}(Li)$_{\odot}$ = 1.05\,dex \citep{Asplund_2009}. Synthetic spectra were computed using the IDL/Fortran-based Spectrum Investigation Utility \citep[SIU;][]{Reetz1999PhDT.......216R} with 1D LTE MARCS model atmospheres \citep{Gustafsson2008}.

The atomic data around the \ion{Li}{1} resonance line and the atomic model of Li were adopted from \citet{Shi2007}. The detailed schematic of the measurement procedure is provided in Figure~\ref{fig:ali_meas} in the Appendix.


\subsection{The validation of our results} \label{subsec:general}

\begin{figure}
    \centering
    \includegraphics[width=1\linewidth]{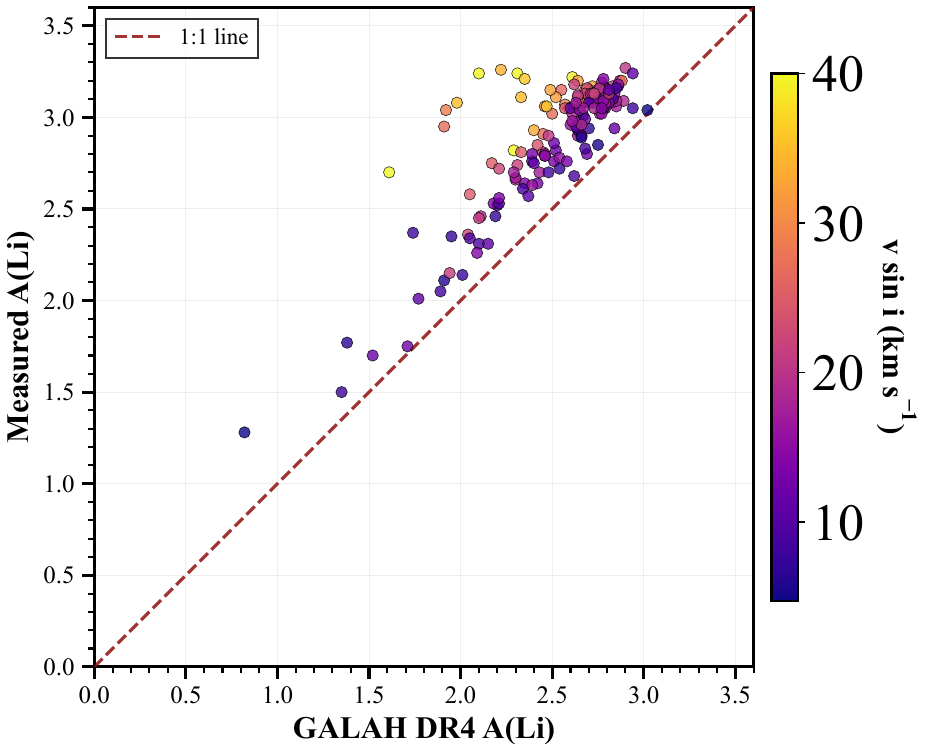}
    \caption{The comparison of our measured lithium abundances with those from \textsc{galah} DR4. Our values are systematically higher than those from \textsc{galah} pipeline, especially for the rapidly rotating stars ($v \sin i$). The color bar is coded by the values of $v \sin i$.}
    \label{fig:fig2}
\end{figure}

The comparison of our measured lithium abundances with those from \textsc{galah} DR4 is shown in Figure~\ref{fig:fig2}, and a systematic offset can be found, i.e., own Li abundances are systematically higher than those from \textsc{galah} pipeline, especially for the rapidly rotating stars ($v \sin i$), Which can likely be due to the higher $v \sin i$ adopted by the automated spectral analysis method. 
Our manual spectral synthesis approach allows for more nuanced local continuum placement and better matching of the broadened line wings. 
This discrepancy rather indicates the limitation of automated abundance determination for high $v \sin i$ stars.

In order to validate the $v \sin i$ values, we independently measured the $v \sin i$ for several sample stars by fitting the line broadening in the spectra using the SIU package. 
We found a good agreement with the values provided in \textsc{GALAH} DR4. The stellar atmospheric parameters and measured lithium abundances of all the 211 sample stars are specified in Table~\ref{tab:your_label} in the Appendix.
 




\section{RESULT} \label{result}

\begin{figure*}
  \centering
  \includegraphics[width=0.9\textwidth]{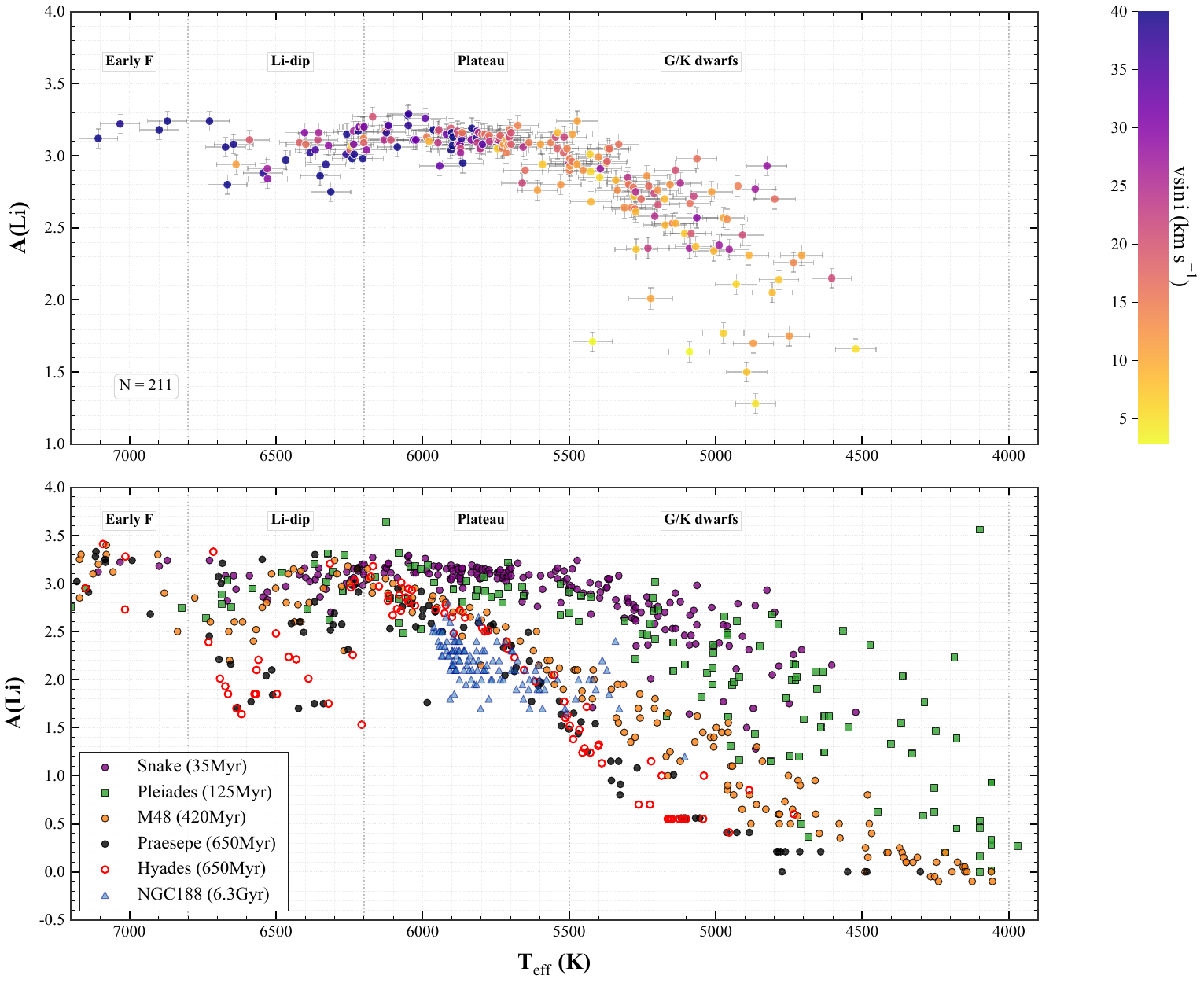} 
\caption{$A\mathrm{(Li)}$ as a function of \tefft. Top: a Li-dip is clearly present among the 211 Snake member stars in the \tefft\ range 6200--6800\,K, with a depletion depth $\Delta A\mathrm{(Li)} \simeq 0.40$\,dex. Points are colour-coded by the values of $v \sin i$; hotter stars generally rotate faster. Error bars represent the typical measurement uncertainty of $A\mathrm{(Li)} = 0.1$\,dex. Bottom: comparison of the $A\mathrm{(Li)}$--$T_{\mathrm{eff}}$ distributions with those of other open clusters: the Pleiades \citep[125\,Myr, solar metallicity;][]{Bouvier2018}, M\,48 \citep[420\,Myr, metal-poor; ][]{Sun_2023}, Praesepe and Hyades \citep[650\,Myr, metal-rich;][]{Cummings_2017}, and NGC\,188 \citep[6.3\,Gyr, solar metallicity;][]{Sun_2025}.}
  \label{fig:lithium} 
\end{figure*}

Figure~\ref{fig:lithium} displays the lithium abundance, {\it A}(Li), as a function of the effective temperature ($T_{\rm eff}$) for the Snake members, and 
we will examine it in detail considering different intervals of $T_{\rm eff}$.    

\subsection{The Li-dip stars}
\label{subsec:LiDip}
Figure~\ref{fig:lithium} clearly shows a Li-dip in the $T_{\rm eff}$ range of 6200-6800\,K for Snake members. The abundance of lithium is significantly lower ({\it A}(Li) $\simeq$ 2.80\,dex) at the dip center near 6500\,K, representing approximately 0.40\,dex depletion relative to the primordial abundance level. A significant abundance dispersion is present in this temperature range, with 35 members of the cluster members spanning {\it A}(Li) from 2.75\,dex to 3.24\,dex, although these stars have a homogeneous metallicity. 
We noted that the lithium abundances provided by \textsc{galah} DR4 even give a deeper Li dip phenomenon in the 6500\,K temperature range. 

\begin{table}[htbp]
\centering
\caption{The median Li abundances for stars in the Li-dip region with high and low $v\sin i$ and different \tefft\ bins}
\label{tab:lidip_rotation_analysis}
\begin{tabular}{ccccccc}
\toprule
\tefft & \multicolumn{3}{c}{$v\sin i > 25$ km/s} & \multicolumn{3}{c}{$v\sin i \leq 25$ km/s} \\
\cmidrule(lr){2-4} \cmidrule(lr){5-7}
 & $N$ & $\langle$A(Li)$\rangle$ & $\sigma/\sqrt{N}$ & $N$ & $\langle$A(Li)$\rangle$ & $\sigma/\sqrt{N}$ \\
(K) & & \multicolumn{2}{c}{(dex)} &  & \multicolumn{2}{c}{(dex)}  \\
\midrule
6200--6350 & 13 & 3.01 & 0.04 & 5 & 3.09 & 0.03 \\
6350--6500 & 5 & 3.04 & 0.04 & 3 & 3.09 & 0.01 \\
6500--6650 & 4 & 2.90 & 0.05 & 2 & 3.03 & 0.09 \\
6650--6800 & 3 & 3.06 & 0.13 & 0 & -- & -- \\
\bottomrule
\end{tabular}
\end{table}

Additionally, a clear anti-correlation between $v\sin i$ and \textit{A}(Li) is evident in Figure~\ref{fig:lithium}, where stars with higher $v\sin i$ systematically exhibit lower \textit{A}(Li). To quantify this trend within the Li-dip region, we present in Table~\ref{tab:lidip_rotation_analysis} the median \textit{A}(Li) values binned by \tefft\ (150\,K intervals) and categorized by rotation: fast rotators ($v\sin i > 25$ km~s$^{-1}$) and slow rotators ($v\sin i \leq 25$ km~s$^{-1}$). The associated uncertainty for each bin is given by $\sigma/\sqrt{N}$, where $\sigma$ and $N$ are the standard deviations of \textit{A}(Li) and the number of stars in the bin, respectively. Across the Li-dip, fast rotators have a mean abundance of $\langle \textit{A}\mathrm{(Li)}\rangle = 3.02 \pm 0.03$\,dex, while slow rotators show a slightly higher mean of $\langle \textit{A}\mathrm{(Li)}\rangle = 3.08 \pm 0.02$\,dex. This systematic difference confirms the negative correlation and exceeds the typical measurement uncertainties of the lithium abundances.


\subsection{Early F-type Dwarfs and Li Plateau stars}

As shown in Figure~\ref{fig:lithium}, the early F-type ($T_{\rm eff} \sim$ 6800\,K -- 7200\,K) and lithium plateau (5500\,K -- 6200\,K) stars, their lithium abundances exhibit remarkable uniformity with a mean value of {\it A}(Li) 3.19$\pm$0.04\,dex and 3.10$\pm$0.07\,dex, respectively (see Figure~\ref{fig:lithium}). This plateau indicates that stars with this $T_{\rm eff}$ region possess sufficient shallow convection zone to prevent significant lithium destruction during their pre-main-sequence evolution. Therefore, these stars contain a Li plateau, a region of relatively constant Li abundance.  



\subsection{ Late G- and K-type Dwarfs}

For late G- and K-type dwarf stars, as usual, their Li abundance undergoes rapid depletion and decreases with decreasing effective temperature. This phenomenon is in agreement with the ``standard” stellar evolution models.


\section{Discussions}
\label{sec:discussion}
This study reports the detection of a pronounced Li-dip in the temperature range of 6200 to 6800\,K (F-type stars) of the open cluster Snake (age = $35 \pm 5$\,Myr). This result suggests that the Li-dip can appear at an age as early as 35\,Myr, and is approximately 100\,Myr earlier than the previous observations. We also note that within the dip temperature range, a significant correlation can be seen between rotational velocity and lithium depletion, i.e., the fast rotators ($v$ sin$i > 25$\,km\,s$^{-1}$) exhibit stronger lithium depletion than the slow ones ($v$ sin$i < 25$\,km\,s$^{-1}$). \citet{Charbonneau1988ApJ} predicted that for stars with $T_{\rm eff}<$ 7000\,K and rotational velocities larger than 25 km s$^{-1}$ of the age of Hyades, the meridional circulation is rapid enough to bring the surface matter to deep enough depleting of its Li through nuclear burning, and an under-abundance of Li is then expected. Very recently, \citet{Li2025ApJ...991..149L} also noted that rotational energy governs the efficiency of meridional circulation, and a higher angular momentum strengthens large-scale circulation cells, thereby increasing the rate of lithium transport to the destruction zone.

For the early F-type stars with a temperature range of 6800--7200\,K, they exhibit relatively rapid rotation, resulting in a weaker \ion{Li}{1} resonance line at 6708\,\AA, which leads to a large uncertainty of the derived lithium abundance. However, current data suggest that there is no obvious lithium depletion of stars within this temperature interval. 

A similar phenomenon can be seen for stars in the lithium plateau region with a temperature range of 5500--6200\,K, where a distinct lithium plateau is clearly observed (see Figure~\ref{fig:lithium}). \citet{Sun_2023} suggested that this plateau is within a \tefft\ range of 6000--6200\,K, and \citet{Sun_2025} noted that there are even no plateau stars in the very old cluster NGC\,188 ($\sim$6.3\,Gyr). This indicates that the effective temperature range of Li plateau may depend on the age. In order to investigate this phenomenon, the lower panel of Figure~\ref{fig:lithium} compares A(Li) as a function of $T_{\rm eff}$ in the Snake with that in five open clusters of different ages. A lower temperature edge of the lithium plateau can be found for young clusters, such as Snake and Pleiades; this result suggests that the temperature range of Li plateau depends on age, and the lower temperature edge can reach as low as 5500\,K for the young Snake. This will be very helpful for us to understand the Li depletion mechanisms in young stellar populations.

For late G- and K-type dwarfs ($< 5500$\,K), as their  effective temperature decreases, the depth of the convection zone increases significantly. Therefore, the temperature of the base convection zone
sufficient for lithium depletion, and the depletion will increase with increasing depth of the convection zone, 
This leads to a clear anti-correlation between lithium abundance and effective temperature for the late G- and K-dwarfs, where lithium abundance decreases markedly with the decreasing stellar temperature.


\section{Conclusion} 
\label{sec:conclusion}

Based on the high-resolution spectra ($R \sim 28,000$) from \textsc{galah} DR4, we find a Li-dip of 0.40\,dex in the 35$\pm$5\,Myr Snake members with $T_{\rm eff}$ between 6200--6800 \,K. The detection of the Li-dip in the Snake revolutionizes our understanding of early stellar Li depletion. Previous observations found that Li-dip onsets around $\gtrsim$150\,Myr.

Also, a prominent anti-correlation between rotation velocity and Li abundance is seen with stars of higher projected rotational velocities ($v\sin i$) showing systematically lower Li abundances.

We noted that the lower temperature edge of the Li plateau can reach as low as 5500\,K for the young Snake, suggesting an age-dependent broadening of the plateau, which has not been taken into account in previous studies. 

The meridional circulation of rapidly rotating F-type stars can transport lithium to the deep convection zone with $T >2.5 \times 10^6$\,K; this can lead to the occurrence of the Li-dip.

Future observations will leverage large-scale Li catalogs, such as \textsc{galah} \citep{Buder2025PASA}, LAMOST \citep{Ding_2024}, 4MOST \citep{Storm2025arXiv251215888S}, and WEAVE \citep{Jin2024MNRAS.530.2688J} to map Li--$T_{\rm eff}$ relationships across diverse environments, resolving the roles of cluster density, metallicity spreads, and accretion anomalies. 

\begin{acknowledgments}

This work is supported by the Strategic Priority Research Program of Chinese Academy of Sciences, grant No.\,XDB1160101. This research is also supported by the National Natural Science Foundation of China under Grant Nos.\,12373033, 12090040/4, 12373036, 12222305, 12427804, the National Key R$\rm\& $D Program of China 2024YFA1611903, the Scientific Instrument Developing Project of the Chinese Academy of Sciences, Grant No.\,ZDKYYQ20220009, the Key Project of Zhejiang Provincial Natural Science Foundation (No.\,ZCLZ25A0301), and the International Partnership Program of the Chinese Academy of Sciences, Grant No.\,178GJHZ2022047GC. 


\end{acknowledgments}

\bibliography{sample7}{}
\bibliographystyle{aasjournalv7}



\appendix
\section{SUPPLEMENTARY DATA AND SPECTRAL ANALYSIS}
\label{app:stellar_params}

\setcounter{figure}{0}  
\setcounter{table}{0}   
\renewcommand{\thefigure}{A.\arabic{figure}}  
\renewcommand{\thetable}{A.\arabic{table}}    

\begin{figure}[htbp]
    \centering
    \includegraphics[width=0.75\textwidth]{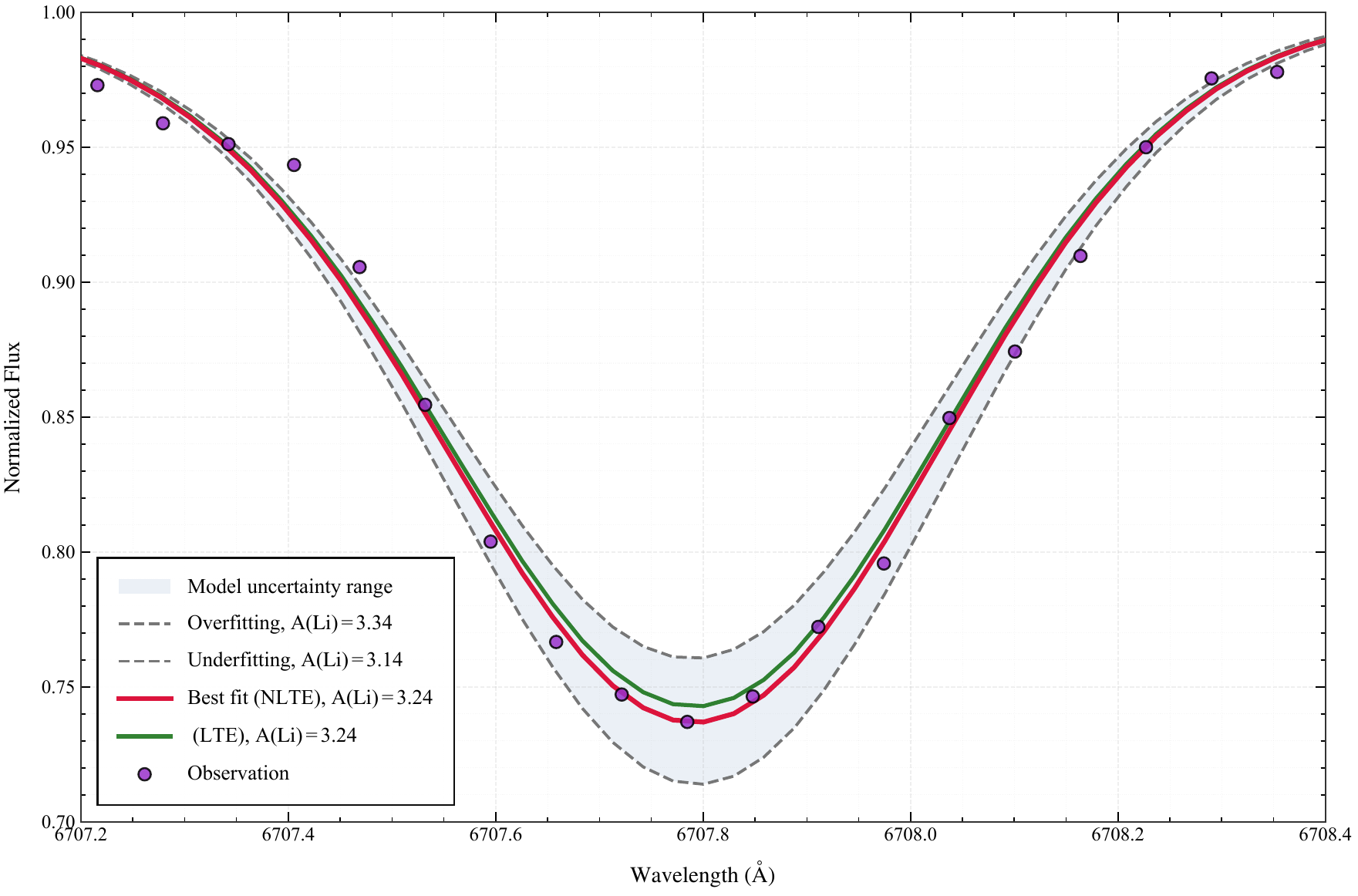}  
    \caption{An example of Li abundances determination. Purple points are the observed spectra, and the red line is the best NLTE fitting synthetic spectra which {\it A}(Li) is 3.24\,dex, and the LTE fitting of same {\it A}(Li) is also provided with the green line. The gray dashed line is under-estimated Li abundance with {\it A}(Li)=3.14\,dex, while the gray dotted line is over-estimated Li abundance with {\it A}(Li) = 3.34\,dex.}
    \label{fig:ali_meas}
\end{figure}

\begingroup
\scriptsize
\setlength{\tabcolsep}{3pt}
\begin{longtable}{@{}lccccccccccc@{}}
\caption{The stellar atmospheric parameters and lithium abundances of the Snake stars}\label{tab:your_label} \\
\toprule
sobject\_id & RA  & Dec & $T_{\mathrm{eff}}$ & $v\sin i$& $\xi_{t}$ & $\log g$ & [Fe/H] & Li$_{\rm NLTE}$ & Li$_{\rm LTE}$ & $M_G$ & bp--rp \\
 & \multicolumn{2}{c}{($^\circ$)} & (K) & \multicolumn{2}{c}{(km\,s$^{-1}$)} & \multicolumn{4}{c}{(dex)} & \multicolumn{2}{c}{(mag)} \\
\midrule
\endfirsthead

\toprule
\multicolumn{12}{c}{\small\textit{Continued from Previous Page}} \\
\midrule
sobject\_id & RA  & Dec & $T_{\mathrm{eff}}$ & $v\sin i$& $\xi_{t}$ & $\log g$ & [Fe/H] & Li$_{\rm NLTE}$ & Li$_{\rm LTE}$ & $M_G$ & bp--rp \\
 & \multicolumn{2}{c}{($^\circ$)} & (K) & \multicolumn{2}{c}{(km\,s$^{-1}$)} & \multicolumn{4}{c}{(dex)} & \multicolumn{2}{c}{(mag)} \\
\midrule
\endhead

\midrule
\multicolumn{12}{r}{\small\textit{Continued on next page}} \\
\endfoot

\bottomrule
\endfoot

\bottomrule
\multicolumn{12}{c}{\footnotesize 
\begin{minipage}{0.9\textwidth}
\centering
Note: The stellar atmospheric parameters and lithium abundances (both NLTE and LTE) listed here are restricted to stars in the Li-dip (\tefft\ is from 6200\,K to 6800\,K). The complete catalog for the entire sample is available online.
\end{minipage}} \\
\endlastfoot
171031003301352 & 94.401703 & -18.214255 & 7106 & 39.98 & 2.18 & 4.22 & -0.33 & 3.12 & 3.15 & 10.686 & 0.427 \\
171031003301348 & 94.535890 & -18.026028 & 7032 & 39.26 & 1.61 & 4.21 & -0.23 & 3.22 & 3.25 & 10.658 & 0.469 \\
180126002101065 & 97.055649 & -4.833379 & 6898 & 40.00 & 0.40 & 4.29 & -0.17 & 3.18 & 3.19 & 10.714 & 0.529 \\
161011004001046 & 106.614544 & -34.516113 & 6871 & 40.00 & 0.26 & 4.24 & -0.31 & 3.24 & 3.24 & 10.691 & 0.554 \\
221221002601387 & 116.838395 & -38.021476 & 6727 & 40.00 & 0.68 & 4.26 & -0.26 & 3.24 & 3.25 & 11.823 & 0.951 \\
170121002801121 & 79.281712 & 19.297638 & 6672 & 39.63 & 1.39 & 4.20 & -0.17 & 3.06 & 3.07 & 11.339 & 0.844 \\
161011004001151 & 105.288067 & -34.691332 & 6666 & 40.00 & 0.44 & 4.23 & -0.13 & 2.80 & 2.84 & 10.010 & 0.563 \\
220422002101060 & 132.298755 & -42.906893 & 6645 & 39.52 & 0.39 & 4.23 & -0.12 & 3.08 & 3.11 & 11.610 & 0.606 \\
220422002101356 & 132.312369 & -42.229216 & 6636 & 11.21 & 1.62 & 4.27 & -0.01 & 2.94 & 2.96 & 11.444 & 0.556 \\
220422002101233 & 130.846286 & -42.029906 & 6590 & 19.11 & 1.44 & 4.28 & -0.07 & 3.11 & 3.13 & 11.508 & 0.561 \\
221221002601045 & 116.557948 & -38.278258 & 6544 & 38.70 & 0.37 & 4.18 & -0.19 & 2.88 & 2.88 & 11.368 & 0.700 \\
220422002101316 & 131.967846 & -42.311281 & 6530 & 27.14 & 1.53 & 4.21 & -0.04 & 2.91 & 2.95 & 11.694 & 0.590 \\
220422002101338 & 132.403032 & -41.975553 & 6529 & 29.38 & 0.32 & 4.32 & -0.22 & 2.84 & 2.87 & 11.829 & 0.607 \\
220422002101330 & 132.206068 & -42.072085 & 6467 & 40.00 & 0.80 & 4.31 & 0.02 & 2.97 & 3.02 & 11.724 & 0.605 \\
221221002601114 & 116.227316 & -38.318344 & 6420 & 18.45 & 1.14 & 4.33 & -0.06 & 3.09 & 3.12 & 11.685 & 0.701 \\
220422002101105 & 131.883722 & -42.654800 & 6402 & 29.17 & 1.30 & 4.33 & 0.01 & 3.16 & 3.19 & 11.873 & 0.633 \\
230214002201222 & 110.192157 & -31.824701 & 6398 & 16.78 & 1.20 & 4.28 & 0.01 & 3.08 & 3.10 & 11.638 & 0.644 \\
220422002101017 & 132.527096 & -42.625461 & 6384 & 38.92 & 0.71 & 4.17 & -0.03 & 3.02 & 3.03 & 11.867 & 0.649 \\
221225003601144 & 142.258198 & -52.490960 & 6366 & 35.62 & 1.25 & 4.17 & -0.06 & 3.04 & 3.07 & 12.011 & 0.655 \\
230214002201223 & 109.742009 & -31.636219 & 6357 & 21.20 & 1.13 & 4.35 & -0.01 & 3.11 & 3.15 & 11.636 & 0.643 \\
220422002101089 & 131.943791 & -42.668023 & 6354 & 26.34 & 0.35 & 4.28 & -0.09 & 3.16 & 3.17 & 12.083 & 0.726 \\
220422002101057 & 132.557249 & -43.099232 & 6350 & 40.00 & 0.33 & 4.24 & -0.29 & 2.86 & 2.86 & 11.853 & 0.653 \\
220422002101093 & 131.972853 & -43.056199 & 6330 & 39.49 & 0.75 & 4.28 & -0.01 & 2.94 & 2.99 & 12.003 & 0.689 \\
220422002101171 & 130.746132 & -42.893779 & 6321 & 28.77 & 1.14 & 4.33 & -0.03 & 3.07 & 3.08 & 11.893 & 0.649 \\
220422002101202 & 130.986221 & -42.467710 & 6313 & 40.00 & 0.26 & 4.20 & -0.35 & 2.75 & 2.73 & 11.951 & 0.674 \\
180126002101095 & 96.810194 & -5.302163 & 6261 & 38.19 & 1.16 & 4.27 & -0.05 & 3.01 & 3.05 & 11.262 & 0.644 \\
221221002601286 & 116.108052 & -37.361215 & 6259 & 39.55 & 0.88 & 4.27 & -0.08 & 3.15 & 3.19 & 11.771 & 0.726 \\
151109003601115 & 101.189642 & -26.245732 & 6247 & 19.87 & 1.24 & 4.28 & -0.03 & 3.04 & 3.07 & 11.613 & 0.624 \\
220422002101003 & 132.674504 & -42.530623 & 6243 & 4.74 & 1.21 & 4.39 & 0.09 & 3.07 & 3.09 & 11.963 & 0.686 \\
220422001601138 & 109.116389 & -36.801194 & 6240 & 40.00 & 0.47 & 4.31 & -0.12 & 2.98 & 3.00 & 11.208 & 0.658 \\
220422002101182 & 131.218418 & -42.634192 & 6235 & 26.27 & 1.30 & 4.34 & -0.00 & 3.17 & 3.18 & 12.118 & 0.684 \\
230214002201275 & 110.563165 & -31.504564 & 6235 & 30.51 & 1.00 & 4.35 & -0.09 & 3.08 & 3.12 & 11.820 & 0.649 \\
220422002101062 & 132.566291 & -43.234419 & 6233 & 39.33 & 0.78 & 4.34 & 0.03 & 3.01 & 3.02 & 11.941 & 0.675 \\
161011004001273 & 105.811365 & -33.635644 & 6220 & 39.66 & 0.72 & 4.26 & -0.12 & 3.17 & 3.19 & 10.948 & 0.640 \\
220422001601303 & 109.335288 & -36.201915 & 6216 & 32.35 & 1.17 & 4.32 & 0.02 & 3.20 & 3.24 & 11.378 & 0.719 \\
220422002101074 & 132.046116 & -42.767893 & 6206 & 16.69 & 1.09 & 4.32 & 0.05 & 3.20 & 3.21 & 12.179 & 0.697 \\
221221002601165 & 115.542395 & -38.446720 & 6204 & 39.85 & 0.28 & 4.32 & -0.17 & 2.98 & 3.02 & 11.831 & 0.724 \\
220422002101170 & 131.366595 & -42.698656 & 6201 & 13.98 & 0.26 & 4.32 & -0.08 & 3.12 & 3.16 & 12.297 & 0.804 \\
221221002601198 & 115.685913 & -38.103934 & 6201 & 18.17 & 1.29 & 4.19 & -0.06 & 3.09 & 3.11 & 11.674 & 0.763 \\
220422002101249 & 131.118854 & -41.908177 & 6199 & 31.87 & 1.15 & 4.36 & 0.00 & 3.20 & 3.21 & 12.170 & 0.685 \\
220422002101058 & 132.050709 & -42.648388 & 6191 & 28.76 & 0.94 & 4.38 & -0.01 & 3.04 & 3.06 & 12.197 & 0.699 \\
220422002101179 & 131.303981 & -42.643995 & 6170 & 20.37 & 1.11 & 4.39 & 0.07 & 3.27 & 3.27 & 12.248 & 0.701 \\
220123002201219 & 110.390427 & -31.873092 & 6131 & 24.98 & 1.11 & 4.29 & -0.04 & 3.11 & 3.14 & 11.946 & 0.703 \\
221221002601043 & 116.893527 & -38.483054 & 6127 & 14.80 & 1.07 & 4.23 & 0.02 & 3.17 & 3.21 & 12.119 & 0.839 \\
230310002101193 & 139.643514 & -48.080372 & 6123 & 39.25 & 0.85 & 4.30 & 0.06 & 3.16 & 3.20 & 12.574 & 0.754 \\
180126002101142 & 96.264874 & -4.938739 & 6117 & 28.02 & 1.09 & 4.34 & -0.10 & 3.20 & 3.21 & 11.654 & 0.697 \\
220422001601169 & 108.238505 & -37.027448 & 6116 & 35.78 & 0.89 & 4.34 & -0.04 & 3.21 & 3.24 & 11.377 & 0.685 \\
161011004001075 & 106.431066 & -34.862200 & 6108 & 20.63 & 0.88 & 4.38 & 0.01 & 3.11 & 3.15 & 11.341 & 0.696 \\
220422002101088 & 132.037066 & -43.004268 & 6086 & 39.92 & 0.30 & 4.09 & -0.14 & 3.06 & 3.07 & 11.579 & 0.743 \\
221221002601112 & 116.259124 & -38.430297 & 6050 & 39.42 & 0.73 & 4.35 & -0.02 & 3.28 & 3.32 & 11.989 & 0.791 \\
160130004101259 & 106.763579 & -34.882385 & 6049 & 37.78 & 1.17 & 4.35 & -0.04 & 3.21 & 3.21 & 12.213 & 0.746 \\
220422001601392 & 110.234515 & -36.511952 & 6048 & 36.99 & 1.14 & 4.39 & -0.01 & 3.29 & 3.32 & 11.613 & 0.751 \\
220422002101276 & 131.558407 & -41.847733 & 6030 & 34.76 & 1.25 & 4.38 & -0.04 & 3.11 & 3.14 & 12.434 & 0.731 \\
221221002601204 & 115.752374 & -37.958333 & 6023 & 31.89 & 1.19 & 4.36 & -0.09 & 3.11 & 3.16 & 12.163 & 0.830 \\
220422001601354 & 110.178976 & -35.985772 & 5991 & 33.76 & 1.22 & 4.37 & 0.01 & 3.26 & 3.27 & 11.878 & 0.779 \\
220422002101245 & 131.691135 & -42.258128 & 5986 & 15.08 & 0.96 & 4.35 & -0.02 & 3.13 & 3.13 & 12.413 & 0.731 \\
230214002201270 & 110.463518 & -31.432629 & 5978 & 9.07 & 0.95 & 4.43 & -0.05 & 3.10 & 3.15 & 12.249 & 0.738 \\
151231003201298 & 118.187409 & -48.921886 & 5963 & 39.89 & 0.96 & 4.31 & 0.13 & 3.18 & 3.22 & 11.746 & 0.812 \\
220422002101335 & 132.299121 & -42.029643 & 5947 & 14.29 & 0.94 & 4.42 & 0.02 & 3.17 & 3.22 & 12.585 & 0.753 \\
221221002601225 & 115.782600 & -37.925049 & 5947 & 23.17 & 0.93 & 4.40 & -0.06 & 3.09 & 3.10 & 12.382 & 0.870 \\
160123001601395 & 110.825202 & -39.056824 & 5945 & 21.80 & 0.95 & 4.38 & 0.03 & 3.18 & 3.20 & 12.356 & 1.000 \\
220422002101379 & 133.067451 & -42.205812 & 5942 & 33.19 & 1.27 & 4.33 & 0.03 & 2.93 & 2.95 & 12.445 & 0.758 \\
220422002101340 & 132.161628 & -42.237607 & 5919 & 32.49 & 1.53 & 4.26 & -0.05 & 3.15 & 3.19 & 12.447 & 0.808 \\
220422002101098 & 131.917184 & -42.997312 & 5906 & 7.94 & 1.19 & 4.35 & 0.05 & 3.05 & 3.09 & 12.601 & 0.785 \\
210113002601035 & 135.092353 & -48.279206 & 5903 & 40.00 & 0.25 & 4.24 & -0.26 & 3.04 & 3.04 & 12.523 & 0.792 \\
180126002101128 & 96.595890 & -4.656262 & 5903 & 19.18 & 1.26 & 4.32 & -0.06 & 3.19 & 3.22 & 11.757 & 0.746 \\
220420001601317 & 117.423556 & -45.656035 & 5901 & 39.97 & 0.69 & 4.32 & -0.09 & 3.16 & 3.19 & 12.177 & 0.797 \\
220422002101380 & 132.124075 & -42.452036 & 5898 & 40.00 & 0.39 & 4.22 & -0.08 & 3.07 & 3.10 & 12.485 & 0.780 \\
170106003601138 & 94.436064 & -17.424929 & 5896 & 39.93 & 0.75 & 4.26 & -0.12 & 3.13 & 3.17 & 11.895 & 0.760 \\
180126002101080 & 96.941724 & -4.772886 & 5881 & 21.89 & 1.49 & 4.41 & 0.02 & 3.17 & 3.19 & 12.067 & 0.810 \\
230214002201028 & 111.332071 & -32.208295 & 5879 & 20.13 & 1.16 & 4.37 & -0.07 & 3.08 & 3.11 & 12.160 & 0.734 \\
220422002101142 & 131.264816 & -43.070360 & 5878 & 20.38 & 1.25 & 4.32 & 0.05 & 3.15 & 3.19 & 12.628 & 0.791 \\
161011004001289 & 105.965100 & -33.680414 & 5872 & 31.53 & 1.25 & 4.32 & -0.12 & 3.08 & 3.09 & 11.587 & 0.730 \\
220422001601255 & 108.856513 & -36.164790 & 5870 & 20.47 & 1.09 & 4.38 & -0.10 & 3.05 & 3.07 & 11.861 & 0.785 \\
150211002701031 & 99.558532 & -21.092731 & 5870 & 26.75 & 1.25 & 4.39 & -0.13 & 3.02 & 3.05 & 12.300 & 0.785 \\
180126002101219 & 95.991282 & -4.088319 & 5867 & 38.02 & 1.21 & 4.26 & -0.18 & 3.12 & 3.15 & 11.577 & 0.712 \\
220422002101236 & 130.897933 & -42.052670 & 5865 & 30.54 & 1.25 & 4.37 & -0.00 & 3.15 & 3.16 & 12.506 & 0.766 \\
221221002601275 & 115.996110 & -37.581304 & 5865 & 14.68 & 1.04 & 4.41 & 0.01 & 3.16 & 3.19 & 12.252 & 0.858 \\
220422002101308 & 131.978818 & -41.984973 & 5863 & 39.98 & 0.29 & 4.30 & -0.20 & 2.95 & 2.99 & 12.506 & 0.786 \\
220422002101068 & 132.427370 & -43.222956 & 5834 & 29.94 & 1.12 & 2.94 & 0.11 & 3.11 & 3.15 & 12.632 & 0.940 \\
221221002601341 & 116.261061 & -38.031161 & 5832 & 40.00 & 0.88 & 4.29 & -0.06 & 3.19 & 3.21 & 12.209 & 0.889 \\
221221002601081 & 116.472084 & -38.575186 & 5817 & 38.95 & 1.18 & 4.34 & -0.08 & 3.12 & 3.14 & 12.419 & 0.870 \\
221221002601216 & 116.265556 & -38.102974 & 5814 & 35.40 & 1.41 & 4.35 & -0.08 & 3.17 & 3.21 & 12.611 & 0.940 \\
221221002601060 & 116.869575 & -38.727589 & 5813 & 22.55 & 1.31 & 4.39 & -0.06 & 3.09 & 3.10 & 12.760 & 1.022 \\
220422001601355 & 110.000399 & -36.114813 & 5809 & 23.39 & 1.13 & 4.40 & -0.09 & 3.16 & 3.19 & 11.973 & 0.811 \\
230214002201214 & 110.660983 & -31.971889 & 5805 & 17.03 & 1.32 & 4.43 & -0.05 & 3.05 & 3.07 & 12.521 & 0.812 \\
170106003601359 & 95.362505 & -16.511094 & 5803 & 24.67 & 1.34 & 4.31 & -0.15 & 3.05 & 3.05 & 12.458 & 0.810 \\
220422002101078 & 132.077892 & -42.845622 & 5797 & 17.96 & 1.24 & 4.42 & -0.00 & 3.15 & 3.16 & 12.782 & 0.811 \\
150209002201287 & 96.123290 & -19.869023 & 5795 & 34.63 & 1.63 & 4.41 & -0.07 & 3.08 & 3.12 & 12.408 & 0.824 \\
140118003001357 & 93.414559 & 5.392336 & 5787 & 17.12 & 1.47 & 4.43 & -0.10 & 3.15 & 3.16 & 12.269 & 0.793 \\
220422001601329 & 109.632779 & -36.062851 & 5778 & 21.76 & 1.29 & 4.40 & -0.02 & 3.13 & 3.15 & 11.977 & 0.813 \\
210113002601184 & 133.756735 & -48.120675 & 5773 & 16.23 & 1.28 & 4.17 & -0.01 & 3.14 & 3.17 & 12.081 & 0.822 \\
221221002601273 & 116.121888 & -37.778947 & 5772 & 23.63 & 1.49 & 4.36 & -0.05 & 3.10 & 3.14 & 12.576 & 0.908 \\
220422002101066 & 132.313915 & -43.039345 & 5746 & 17.58 & 1.38 & 4.39 & -0.04 & 3.11 & 3.13 & 12.883 & 0.849 \\
220422001601046 & 109.936540 & -37.078358 & 5745 & 6.50 & 1.02 & 4.49 & -0.05 & 3.05 & 3.06 & 13.031 & 0.880 \\
221221002601160 & 115.941449 & -38.263361 & 5741 & 33.45 & 1.46 & 4.37 & -0.04 & 3.11 & 3.15 & 12.580 & 0.946 \\
230214002201314 & 110.884144 & -31.702274 & 5737 & 19.58 & 1.23 & 4.45 & -0.03 & 3.12 & 3.16 & 12.562 & 0.824 \\
230214002201022 & 110.980072 & -32.066400 & 5735 & 17.77 & 1.41 & 4.48 & -0.01 & 3.14 & 3.17 & 12.825 & 0.873 \\
220422002101123 & 131.765332 & -42.749125 & 5727 & 35.88 & 0.36 & 4.25 & -0.09 & 3.06 & 3.07 & 12.740 & 0.970 \\
230403001601265 & 123.507220 & -41.617058 & 5727 & 14.34 & 0.79 & 4.18 & 0.00 & 3.05 & 3.08 & 12.736 & 0.821 \\
221221002601178 & 115.981360 & -38.186529 & 5724 & 10.79 & 1.11 & 4.45 & -0.07 & 3.08 & 3.11 & 12.833 & 0.948 \\
220422001601310 & 109.354369 & -36.345695 & 5715 & 14.71 & 1.20 & 4.42 & -0.08 & 3.02 & 3.06 & 12.147 & 0.849 \\
230411000101384 & 144.465128 & -47.847805 & 5715 & 9.95 & 1.04 & 4.44 & 0.06 & 3.11 & 3.15 & 13.363 & 0.890 \\
171031003301337 & 94.421128 & -17.890432 & 5711 & 13.09 & 0.93 & 4.42 & 0.02 & 3.07 & 3.12 & 12.298 & 0.824 \\
220422001601307 & 109.422314 & -35.820750 & 5707 & 13.56 & 1.19 & 4.44 & 0.03 & 3.08 & 3.12 & 12.296 & 0.881 \\
220422002101011 & 132.428561 & -42.577480 & 5706 & 20.86 & 1.47 & 4.38 & -0.02 & 3.13 & 3.14 & 13.030 & 0.859 \\
220422002101359 & 132.118734 & -42.387407 & 5704 & 20.48 & 1.49 & 4.45 & 0.08 & 3.13 & 3.16 & 13.011 & 0.866 \\
220422001601006 & 109.803668 & -36.635754 & 5703 & 15.10 & 1.23 & 4.46 & 0.08 & 3.18 & 3.20 & 12.142 & 0.852 \\
220422002101040 & 132.622039 & -42.886836 & 5698 & 18.95 & 1.38 & 4.41 & -0.02 & 3.08 & 3.09 & 12.895 & 0.837 \\
221221002601183 & 115.375115 & -38.280657 & 5698 & 20.31 & 1.39 & 4.42 & 0.01 & 3.16 & 3.19 & 12.880 & 0.924 \\
220216001601047 & 94.934103 & 5.629696 & 5675 & 14.29 & 1.10 & 4.48 & 0.06 & 3.21 & 3.23 & 12.527 & 0.874 \\
171031003301241 & 93.439998 & -18.326886 & 5660 & 21.47 & 1.20 & 4.29 & -0.24 & 2.81 & 2.85 & 12.272 & 0.779 \\
221221002601044 & 116.474544 & -38.211275 & 5657 & 24.18 & 1.08 & 4.13 & 0.01 & 3.06 & 3.11 & 11.739 & 0.938 \\
220422001601330 & 109.849458 & -35.763501 & 5650 & 17.24 & 1.19 & 4.43 & -0.10 & 2.90 & 2.91 & 12.328 & 0.889 \\
220422002101376 & 132.462440 & -42.347930 & 5637 & 15.11 & 1.27 & 4.43 & 0.04 & 3.09 & 3.14 & 13.052 & 0.854 \\
230214002201083 & 110.944980 & -32.487473 & 5608 & 11.16 & 0.25 & 4.28 & -0.29 & 2.76 & 2.80 & 12.086 & 0.860 \\
161118005201347 & 106.870952 & -36.824589 & 5597 & 10.77 & 1.17 & 4.47 & -0.07 & 3.08 & 3.10 & 12.470 & 0.871 \\
160123001601372 & 110.772893 & -38.832064 & 5591 & 7.31 & 1.26 & 4.52 & 0.04 & 2.94 & 2.96 & 13.699 & 1.125 \\
221221002601299 & 116.300225 & -37.754678 & 5563 & 14.53 & 1.36 & 4.40 & 0.04 & 3.09 & 3.10 & 13.014 & 0.954 \\
181226003601153 & 104.873634 & -36.168942 & 5546 & 21.18 & 1.42 & 4.51 & 0.03 & 3.13 & 3.14 & 12.216 & 0.910 \\
220422002101162 & 131.017736 & -42.916725 & 5540 & 18.36 & 1.22 & 4.43 & -0.05 & 3.05 & 3.09 & 13.195 & 0.892 \\
180129003601017 & 109.438097 & -36.702947 & 5539 & 8.29 & 1.26 & 4.46 & 0.01 & 3.16 & 3.20 & 12.443 & 0.907 \\
180129003601282 & 108.940915 & -36.166852 & 5529 & 12.20 & 0.97 & 4.44 & -0.15 & 2.80 & 2.84 & 12.346 & 0.870 \\
221221002601208 & 115.831435 & -38.093634 & 5519 & 18.78 & 1.32 & 4.35 & -0.03 & 3.02 & 3.06 & 12.984 & 0.997 \\
221221002601032 & 116.549331 & -38.209985 & 5517 & 22.11 & 1.38 & 4.40 & 0.03 & 3.13 & 3.16 & 13.207 & 1.054 \\
220422001601057 & 110.004992 & -37.328811 & 5507 & 14.43 & 1.44 & 4.52 & -0.00 & 3.05 & 3.07 & 12.780 & 0.984 \\
221221002601308 & 116.406598 & -37.507434 & 5501 & 29.43 & 0.82 & 4.34 & -0.10 & 2.95 & 2.96 & 12.850 & 0.983 \\
230404002101039 & 141.728501 & -46.342911 & 5500 & 26.83 & 1.08 & 4.41 & -0.10 & 2.95 & 3.00 & 13.304 & 0.904 \\
221221002601117 & 116.143309 & -38.551306 & 5499 & 14.73 & 1.11 & 4.45 & -0.05 & 2.90 & 2.91 & 13.085 & 0.975 \\
220422002101296 & 131.855374 & -41.786190 & 5497 & 14.81 & 1.22 & 4.44 & -0.02 & 2.98 & 3.00 & 13.159 & 0.887 \\
230214002201355 & 111.150381 & -31.756608 & 5497 & 12.02 & 1.36 & 4.47 & -0.02 & 2.93 & 2.95 & 13.166 & 0.940 \\
220422001601197 & 108.696190 & -36.616360 & 5490 & 16.17 & 1.19 & 4.49 & -0.09 & 2.96 & 3.01 & 12.473 & 0.894 \\
220422001601313 & 109.427691 & -36.156045 & 5489 & 10.79 & 1.49 & 4.43 & -0.00 & 3.15 & 3.16 & 12.803 & 0.985 \\
221221002601202 & 116.202957 & -38.098687 & 5473 & 10.58 & 1.19 & 4.49 & -0.11 & 2.94 & 2.95 & 13.279 & 1.077 \\
220422001601181 & 108.620328 & -36.795923 & 5472 & 10.59 & 1.47 & 4.43 & 0.04 & 3.24 & 3.26 & 12.313 & 0.961 \\
230214002201235 & 110.537630 & -31.854672 & 5452 & 12.58 & 1.39 & 4.38 & -0.05 & 2.90 & 2.95 & 12.497 & 0.919 \\
220422002101015 & 132.337120 & -42.567116 & 5429 & 6.16 & 1.33 & 4.42 & 0.06 & 3.01 & 3.05 & 13.618 & 0.980 \\
181226003601158 & 104.714430 & -36.235895 & 5426 & 5.81 & 0.99 & 4.52 & -0.06 & 2.89 & 2.90 & 12.276 & 0.910 \\
220422002101354 & 132.769680 & -41.946167 & 5426 & 9.08 & 1.18 & 4.52 & -0.07 & 2.68 & 2.71 & 13.425 & 0.920 \\
181225004101044 & 108.202684 & -39.754154 & 5420 & 2.85 & 1.25 & 4.31 & 0.16 & 1.71 & 1.72 & 12.552 & 1.069 \\
220422002101149 & 131.203085 & -43.018406 & 5400 & 11.98 & 1.41 & 4.40 & -0.06 & 2.99 & 3.03 & 13.533 & 0.953 \\
161104004801322 & 96.869653 & -14.788599 & 5396 & 5.65 & 0.84 & 4.22 & -0.10 & 2.85 & 2.87 & 13.210 & 1.041 \\
170106003601353 & 95.378121 & -16.407785 & 5394 & 27.87 & 0.46 & 4.36 & -0.14 & 2.91 & 2.95 & 12.443 & 0.952 \\
220215001601149 & 93.575228 & 3.347864 & 5373 & 11.22 & 1.38 & 4.30 & -0.10 & 2.95 & 2.98 & 12.534 & 0.934 \\
221225003201122 & 122.941396 & -46.260940 & 5371 & 18.62 & 1.50 & 4.41 & -0.09 & 2.96 & 2.99 & 13.156 & 0.961 \\
220422002101323 & 132.225483 & -41.855542 & 5369 & 11.15 & 1.41 & 4.41 & 0.05 & 3.05 & 3.10 & 13.445 & 0.946 \\
180129003601164 & 108.718365 & -36.790449 & 5364 & 14.25 & 1.33 & 4.51 & -0.06 & 3.04 & 3.05 & 12.661 & 0.971 \\
221221002601237 & 116.093890 & -37.882075 & 5363 & 16.85 & 1.51 & 4.42 & -0.10 & 3.05 & 3.07 & 12.974 & 1.001 \\
211217002601385 & 92.601270 & 4.097308 & 5341 & 8.78 & 1.58 & 4.40 & -0.12 & 2.83 & 2.86 & 12.909 & 0.986 \\
221221002601318 & 116.393716 & -37.864693 & 5335 & 13.13 & 1.50 & 4.44 & -0.04 & 2.76 & 2.78 & 13.118 & 1.062 \\
220422001601116 & 108.983233 & -37.544223 & 5331 & 16.43 & 1.71 & 4.46 & -0.07 & 3.08 & 3.09 & 12.659 & 0.956 \\
170106003601125 & 94.557337 & -17.439117 & 5312 & 13.01 & 1.24 & 4.24 & -0.11 & 2.64 & 2.68 & 12.847 & 0.986 \\
161118005201180 & 106.065782 & -36.898259 & 5302 & 11.64 & 1.50 & 4.40 & -0.03 & 2.82 & 2.84 & 12.535 & 1.003 \\
220422002101337 & 132.434411 & -41.907145 & 5300 & 23.92 & 1.49 & 4.50 & -0.10 & 2.85 & 2.88 & 13.625 & 0.985 \\
221221002601226 & 115.851585 & -37.936969 & 5297 & 16.27 & 1.65 & 4.44 & -0.10 & 2.80 & 2.81 & 13.288 & 1.094 \\
220422002101063 & 132.005715 & -42.684866 & 5286 & 14.05 & 1.69 & 4.54 & -0.01 & 2.64 & 2.64 & 13.875 & 1.051 \\
221221002601220 & 115.182896 & -37.795018 & 5281 & 15.93 & 1.46 & 4.45 & -0.12 & 2.78 & 2.80 & 13.211 & 1.026 \\
220422002101051 & 132.351932 & -42.860596 & 5278 & 7.21 & 1.51 & 4.49 & -0.03 & 2.72 & 2.73 & 13.656 & 0.988 \\
221221002601167 & 116.214029 & -38.133704 & 5276 & 15.98 & 1.53 & 4.43 & -0.12 & 2.63 & 2.66 & 13.466 & 1.193 \\
220422002101124 & 131.712964 & -42.792603 & 5273 & 8.96 & 1.45 & 4.63 & -0.04 & 2.61 & 2.63 & 14.000 & 1.065 \\
230214002201046 & 110.944634 & -32.117393 & 5273 & 25.10 & 1.69 & 4.45 & -0.10 & 2.75 & 2.77 & 13.338 & 1.045 \\
220422001601314 & 109.511755 & -35.888755 & 5272 & 7.36 & 1.06 & 4.54 & 0.00 & 2.35 & 2.39 & 13.022 & 1.015 \\
161118005201360 & 106.786406 & -36.809458 & 5255 & 17.02 & 1.83 & 4.45 & -0.05 & 2.70 & 2.70 & 12.619 & 1.001 \\
140118002001218 & 90.510137 & 7.759610 & 5235 & 12.03 & 1.62 & 4.41 & -0.09 & 2.86 & 2.89 & 12.999 & 1.028 \\
221104003101206 & 102.202250 & -23.955566 & 5231 & 22.61 & 0.88 & 4.19 & -0.24 & 2.36 & 2.39 & 12.956 & 1.119 \\
230214002201074 & 110.907277 & -32.225408 & 5229 & 17.07 & 1.70 & 4.46 & -0.06 & 2.79 & 2.83 & 13.358 & 1.031 \\
230209001901195 & 132.287107 & -50.047217 & 5222 & 10.89 & 0.25 & 4.39 & -0.34 & 2.01 & 2.04 & 13.216 & 1.143 \\
221221002601154 & 115.390670 & -38.692459 & 5210 & 21.68 & 1.55 & 4.46 & -0.06 & 2.74 & 2.79 & 13.241 & 1.073 \\
220422002101139 & 131.714057 & -42.689408 & 5208 & 24.67 & 1.60 & 4.51 & -0.07 & 2.58 & 2.61 & 13.858 & 1.072 \\
230214002201142 & 110.355374 & -32.470685 & 5198 & 12.81 & 1.62 & 4.51 & -0.09 & 2.76 & 2.79 & 13.572 & 1.038 \\
221221002601169 & 115.650850 & -38.359551 & 5197 & 20.53 & 1.78 & 4.46 & -0.09 & 2.66 & 2.68 & 13.431 & 1.137 \\
170106003601009 & 95.232478 & -16.945163 & 5174 & 8.55 & 1.27 & 4.47 & -0.05 & 2.70 & 2.71 & 13.637 & 1.139 \\
220422002101261 & 131.288422 & -41.821026 & 5172 & 9.19 & 1.28 & 4.58 & 0.02 & 2.52 & 2.56 & 13.933 & 1.053 \\
220422002101294 & 131.817126 & -41.830692 & 5156 & 12.38 & 1.66 & 4.52 & -0.05 & 2.80 & 2.84 & 13.992 & 1.100 \\
220422002101102 & 131.875699 & -43.113849 & 5147 & 13.85 & 1.73 & 4.58 & -0.02 & 2.53 & 2.55 & 14.207 & 1.123 \\
220422001601020 & 109.852318 & -36.655157 & 5138 & 20.46 & 1.76 & 4.40 & -0.10 & 2.90 & 2.95 & 13.552 & 1.072 \\
220422002101043 & 132.234652 & -42.672102 & 5137 & 9.53 & 1.43 & 4.50 & -0.02 & 2.53 & 2.56 & 13.918 & 1.072 \\
230214002201326 & 111.109710 & -31.405293 & 5121 & 25.05 & 1.64 & 4.45 & -0.04 & 2.81 & 2.85 & 13.471 & 1.092 \\
220422002101280 & 131.556615 & -41.637128 & 5106 & 7.14 & 1.40 & 4.57 & 0.01 & 2.46 & 2.49 & 14.069 & 1.073 \\
230205001601263 & 100.184202 & -5.432203 & 5090 & 2.82 & 1.14 & 4.56 & 0.09 & 1.64 & 1.68 & 12.965 & 1.086 \\
220422002101131 & 131.348465 & -43.269462 & 5090 & 27.00 & 0.39 & 4.40 & -0.19 & 2.36 & 2.40 & 13.787 & 1.071 \\
220422002101314 & 132.041920 & -42.046377 & 5088 & 16.24 & 1.60 & 4.34 & -0.08 & 2.67 & 2.72 & 13.162 & 1.066 \\
220422002101332 & 132.240509 & -42.054310 & 5084 & 19.39 & 1.81 & 4.51 & 0.01 & 2.46 & 2.46 & 13.959 & 1.103 \\
220422002101094 & 131.983556 & -43.258308 & 5075 & 22.16 & 1.77 & 4.51 & -0.03 & 2.72 & 2.76 & 13.866 & 1.069 \\
220422002101215 & 131.870938 & -42.493525 & 5068 & 7.98 & 1.36 & 4.60 & 0.02 & 2.37 & 2.42 & 14.162 & 1.115 \\
220422002101235 & 131.068697 & -42.111738 & 5065 & 29.18 & 0.47 & 4.48 & -0.38 & 2.57 & 2.58 & 13.880 & 1.068 \\
220422001601076 & 109.443298 & -36.991481 & 5064 & 16.41 & 1.57 & 4.49 & -0.07 & 2.98 & 2.99 & 13.018 & 1.050 \\
220422002101033 & 132.348219 & -42.719944 & 5014 & 11.17 & 1.86 & 4.52 & -0.05 & 2.75 & 2.75 & 14.351 & 1.189 \\
220422002101307 & 132.062871 & -41.517919 & 5006 & 9.04 & 1.51 & 3.92 & -0.11 & 2.34 & 2.36 & 14.150 & 1.081 \\
220422002101181 & 131.065649 & -42.675805 & 4988 & 27.37 & 0.28 & 4.49 & -0.27 & 2.38 & 2.42 & 14.144 & 1.179 \\
220422001601281 & 108.986144 & -35.736535 & 4974 & 6.58 & 1.07 & 4.56 & -0.08 & 1.77 & 1.80 & 13.445 & 1.163 \\
220422002101266 & 131.420415 & -41.885073 & 4974 & 10.22 & 1.77 & 4.51 & -0.08 & 2.57 & 2.61 & 14.169 & 1.143 \\
220420001601124 & 117.120578 & -46.265620 & 4961 & 13.22 & 1.55 & 4.45 & -0.09 & 2.56 & 2.57 & 13.053 & 1.198 \\
220422002101048 & 131.996402 & -42.590218 & 4954 & 26.65 & 0.93 & 4.31 & -0.10 & 2.35 & 2.38 & 13.464 & 1.109 \\
230214002201362 & 111.337758 & -31.697954 & 4930 & 5.99 & 1.44 & 4.51 & -0.04 & 2.11 & 2.16 & 13.919 & 1.198 \\
221221002101345 & 88.945064 & 6.514603 & 4924 & 14.85 & 1.63 & 4.46 & 0.09 & 2.79 & 2.83 & 13.466 & 1.237 \\
220422002101234 & 131.546866 & -42.261890 & 4908 & 20.97 & 1.82 & 4.49 & 0.05 & 2.45 & 2.47 & 14.426 & 1.253 \\
220422002101030 & 132.444360 & -42.713624 & 4894 & 7.67 & 1.79 & 4.55 & -0.05 & 1.50 & 1.52 & 14.328 & 1.194 \\
220422002101239 & 131.452042 & -42.248606 & 4887 & 9.31 & 1.35 & 4.48 & -0.09 & 2.31 & 2.32 & 14.223 & 1.147 \\
220422002101120 & 131.664599 & -43.010191 & 4872 & 11.88 & 1.48 & 4.55 & -0.05 & 1.70 & 1.74 & 14.292 & 1.189 \\
230214002201292 & 110.813555 & -31.738505 & 4865 & 27.27 & 1.43 & 4.44 & -0.08 & 2.77 & 2.79 & 13.969 & 1.297 \\
161118005201394 & 106.849497 & -36.962815 & 4864 & 4.80 & 1.03 & 4.62 & -0.17 & 1.28 & 1.32 & 13.655 & 1.183 \\
211214001201082 & 96.455625 & -4.816836 & 4825 & 28.13 & 1.26 & 4.44 & -0.12 & 2.93 & 2.94 & 13.575 & 1.320 \\
220420001601291 & 117.031662 & -45.258084 & 4807 & 8.39 & 1.69 & 4.31 & -0.10 & 2.05 & 2.05 & 13.586 & 1.218 \\
230214002201273 & 110.523339 & -31.451256 & 4799 & 17.19 & 1.57 & 4.48 & -0.05 & 2.70 & 2.74 & 13.936 & 1.279 \\
150204001601237 & 101.685056 & -35.006333 & 4785 & 7.42 & 1.44 & 4.52 & -0.11 & 2.14 & 2.18 & 12.926 & 1.309 \\
220422002101151 & 131.988811 & -42.584522 & 4749 & 12.06 & 1.84 & 4.45 & -0.10 & 1.75 & 1.78 & 14.711 & 1.349 \\
160123001601184 & 109.322130 & -39.233538 & 4735 & 13.86 & 1.60 & 4.49 & -0.09 & 2.26 & 2.27 & 13.205 & 1.317 \\
181226003601065 & 105.756482 & -36.461028 & 4707 & 10.69 & 1.58 & 4.52 & -0.04 & 2.31 & 2.34 & 13.580 & 1.457 \\
211215002601366 & 89.173665 & 8.152234 & 4604 & 21.06 & 1.57 & 4.47 & -0.15 & 2.15 & 2.17 & 13.568 & 1.402 \\
220422002101309 & 131.914686 & -42.042703 & 4522 & 5.09 & 1.11 & 4.48 & -0.19 & 1.66 & 1.68 & 14.878 & 1.474 \\
\end{longtable}
\endgroup

\end{document}